\documentclass[pre,twocolumn,amsmath,amssymb,amsfonts,floatfix,superscriptaddress,showkeys,showpacs]{revtex4-1}
\usepackage{dcolumn}  % Align table columns on decimal point
\usepackage{multirow}
\usepackage{hyperref} 
\usepackage{verbatim}
\usepackage{bm} % bold math 
\usepackage{graphicx}
\usepackage{epstopdf}
\usepackage{ifpdf} % This lets us get figures right both in pdf and ps versions
\usepackage{epsfig}
\usepackage{color}
\usepackage[usenames,dvipsnames]{xcolor}
\usepackage{float}
\usepackage[colorinlistoftodos]{todonotes}

\begin{document}

\title{Effective interactions between inclusions in an active bath}

\author{Mahdi Zaeifi Yamchi}\thanks{mahdi.zaeifi@ipm.ir}
\affiliation{School of Physics, Institute for Research in Fundamental Sciences (IPM), Tehran 19395-5531, Iran}

\author{Ali Naji}\thanks{a.naji@ipm.ir (corresponding author)}
\affiliation{School of Physics, Institute for Research in Fundamental Sciences (IPM), Tehran 19395-5531, Iran}

%\date{\today}

\begin{abstract}
We study effective two- and three-body interactions between non-active colloidal inclusions in an active bath of chiral or non-chiral particles, using  Brownian Dynamics simulations within a standard, two-dimensional  model of disk-shaped inclusions and active particles. In a non-chiral active bath, we first corroborate previous findings on effective two-body repulsion mediated between the inclusions by elucidating the detailed non-monotonic features of the two-body force profiles, including a primary maximum, and a secondary hump at larger separations that was not previously reported. We then show that these features arise directly from the formation, and sequential overlaps, of circular layers (or `rings') of active particles around the inclusions, as the latter are brought to small surface separations. These rings extend to radial distances of a few active-particle radii from the surface of inclusions, giving the hard-core inclusions relatively thick, soft, repulsive `shoulders', whose multiple overlaps then enable significant (non-pairwise) three-body forces in both non-chiral and chiral active baths. The resulting three-body forces can even exceed the two-body forces in magnitude and display distinct repulsive and attractive regimes at intermediate to large self-propulsion strengths. In a chiral active bath, we show that, while active particles still tend to accumulate at the immediate vicinity of the inclusions, they exhibit strong depletion from the intervening region between the inclusions, and partial depletion from relatively thick, circular, zones further away from the inclusions. In this case, the effective, predominantly repulsive, interactions between the inclusions turn to active, chirality-induced, depletion-type attractions, acting over an extended range of separations. 
\end{abstract}

%\pacs{ 64.70. Pf}
% \keywords{Self-propelled particles, chiral active matter, depletion force}

\maketitle

%SECTION%%%%%%%%%%%%%%%%%%%%%%%%%%%%%%%%%%%%%%%%%%%%%%%%%%%%%%
\section{Introduction}
\label{sec:intro}

Active systems comprising self-propelled particles have emerged as a novel class of non-equilibrium systems, exhibiting  
% in which the combination of active particle motion, noise and inter-particle interactions often results in unusual phases with diverse 
unusual phases with diverse structural, dynamical and mechanical properties out of equilibrium  \cite{purcell:1977,berg2003coli,Lauga:RPP009,ramaswamyreview,golestanian_review,laugascallop1,Romanczuk:EPJ2012,Marchetti:RMP2013,Yeomans:EPJ2014,Elgeti:RPP2015,cates_review,Goldstein:Review,Lauga:ANNREVF2016,ZottlStark_review,bechinger_review,Vicsek:PRL1995,Helbing:RMP2001,Simha:PRL2002,Shenoy:PNAS2007,Kudrolli:PRL2008,Drocco:PRE2012}. Typical examples of active particles include asymmetrically coated, or Janus,  micro-/nano-particles that exhibit catalytic surface reactions and, hence, self-propulsion, in specific solvent mixtures  \cite{Walther:2013,Walther:softmatt2008,Jiang:2010,Perro:2005,sanojanus,valadares2010catalytic,ramin07,ramin_molmach,golestanian_njp,Douglass:NatPhot2012,Volpe:SciRep2014,Bianchi:SciRep2016,Buttinoni:PRL2013,Mano:JACS2005}. These synthetic particles have enabled detailed study of different aspects of active processes involving self-propulsion \cite{Walther:2013,Walther:softmatt2008,Jiang:2010,Perro:2005,becker,StarkPRL2014,volpe1,sanojanus,valadares2010catalytic,ramin07,ramin_molmach,najafiramin05,golestanian_njp,najafiramin,Douglass:NatPhot2012,Volpe:SciRep2014,Bianchi:SciRep2016,Buttinoni:PRL2013,Mano:JACS2005,Romanczuk:PRL2011,Grosmann:NJPhys2013,Grosmann:NJPhys2012,Romanczuk:EPJ2015,Popescu1,Popescu2,Popescu3}. Self-propelled particles are also abundant in biology with examples including many of the living organisms, such as multiflagellate bacterium {\em E. coli} \cite{berg2003coli}, biflagellate alga {\em C. reinhardtii} \cite{Goldstein:Review}  and uniflagellated sperm cells  \cite{SHACK1974555,Woolley01082003}, which contain engines of their own to self-propel in fluid media and in the absence of external forces \cite{purcell:1977,berg2003coli,Lauga:RPP009,ramaswamyreview,golestanian_review,laugascallop1,Romanczuk:EPJ2012,Marchetti:RMP2013,Yeomans:EPJ2014,Elgeti:RPP2015,cates_review,Goldstein:Review,Lauga:ANNREVF2016,ZottlStark_review,bechinger_review}. 

In the most common models of self-propelled particles, it is assumed that the particles possess uniaxial symmetry (with examples ranging from Janus micro-/nano-spheres and rods to {\em E. coli})
%(which is a suitable approximation for a wide class of swimmers, including self-propelled micro- and nanorods as well as living organisms such as {\em E. coli}) 
and that the direction of the self-propulsion is along the axis of symmetry of the particle. Such active particles thus tend to orient along the self-propelling  force and move in a straight line, although their translational and rotational motion may also be subject to passive and/or active noise. While passive noise (due, e.g., to  thermal ambient fluctuations) is independent of the direction of  motion, active noise (due, e.g., to fluctuations in the internal self-propulsion mechanism) correlates with that direction \cite{Romanczuk:EPJ2012,Romanczuk:PRL2011,Grosmann:NJPhys2013,Grosmann:NJPhys2012,Romanczuk:EPJ2015}.

In general, asymmetries in self-propulsion mechanism can generate a more complex behavior, including, e.g., situations with an unbalanced torque acting on the particle; in this case, the direction of motion and that of the active force are no longer aligned and the active particles tend to execute circular motion, with a characteristic radius  inversely proportional to the magnitude of the torque (or, equivalently, their intrinsic angular velocity). Such {\em chiral} self-propelled particles %or {\em circle swimmers} 
have attracted mounting interest over the last few years  \cite{Teeffelen:PRE2008,Bao-quan:2015,Reichhardt:2013,Mijalkov:2015,Breier:2014,Xue:EPL2015,Xue:EPJST2014,Romanczuk:EPJ2012,Volpe:2013,Volpe:2014,Crespi:2013,Crespi:PRL2015,Li:PRE2014,Bechinger:PRL2013,Lowen:EPJST2016,Weber:PRE2011,Shelley:2013,Friedrich:2008,Friedrich:2009,Golestanian:PRE2010,Mirkovic:SMLL2010,Boymelgreen:PRE2014,
Marine:PRE2013,Wheat:Langmuir2010,Xue:PNAS2012,DiLeonardo:2011}. Depending on the sign of the torque, one can consider two types of left- and right-handed chiral self-propelled particles. Separation strategies for the two different types of active particles %(e.g., using geometric barriers) 
have been discussed in recent works \cite{Volpe:2013,Volpe:2014,Bao-quan:2015}. In addition, chiral self-propelled particles have shown the potential to be used as carries of non-chiral particles in ratchet channels \cite{Ai:SciRep2016}. Another interesting, and yet unexplored, aspect of chiral self-propelled particles, to be addressed among other problems in this paper, is the non-equilibrium effective (possibly, depletion-type) interactions that an active bath of such particles can mediate between external, non-active, colloidal objects.

Depletion interactions have been studied extensively in the context of {\em non-active} macromolecular mixtures in equilibrium due both to their fundamental importance in the theory of colloidal  stability \cite{Lekkerkerker:2011,Likos:2001} and their numerous technological applications as, e.g., in protein crystallization \cite{Jia:BioPhy2005} and phase separations of 
%colloids and 
colloid/polymer mixtures \cite{Tuinier:ACIS2003}. %{Ye:PRE1996}. 
%, protein crystallization \cite{Jia:BioPhy2005}, and structural assemblies in cell \cite{Marenduzzo:JCB2006}. 
In their most basic form, equilibrium depletion interactions can be established as effective, short-ranged, attractive forces between two hard spheres (inclusions)  immersed in a bath of thermalized hard spheres of smaller size (depletants) \cite{Lekkerkerker:2011,Likos:2001}. In this case, steric repulsions lead to exclusion of small spheres from the immediate vicinity of the inclusions, forming thin depletion layers (of thickness equal to small sphere radius) around them. The overlap between these layers at small surface separations creates a narrow intervening region of complete depletion between the inclusions and, hence, a net attractive force  %(of purely entropic origin) 
between them. 
%Depletion interactions can, in general, be strongly size-, shape- and concentration-dependent. 
% and can be repulsive or attractive \cite{Crocker:PRL1999}. 

Depletion-type interactions have also been studied in non-equilibrium systems involving colloidal inclusions immersed in a bath of {\em active} particles \cite{Likos:2003,Angelani:2011,Schwarz-Linek13032012,Harder:2014,Ni:PRL2015,Leite:PRE2016,Smallenburg:2015,Angelani:2011,Schwarz-Linek13032012,Cacciuto:2016a,Cacciuto:2016b}, and their relation to Casimir-type forces has been discussed \cite{Reichhardt:PRE2014}.  
%These studies include both experiments (e.g., using {\em E. coli} as depletant \cite{Angelani:2011,Schwarz-Linek13032012}) and numerical simulations based on standard models of self-propelled particles to investigate the effect of active bath on the effective interactions between suspended, non-active colloidal particles.
 Studies of the phase behavior of suspensions of non-motile and motile {\em E. coli}, used as non-active and active colloids, respectively, in the presence of non-adsorbing polymers as depletant, show that activeness suppresses the depletion-driven phase separation of {\em E. coli} \cite{Schwarz-Linek13032012}. Harder et al. \cite{Harder:2014} used Brownian Dynamics simulations to study the steady-state, effective two-body interactions mediated between large non-active colloids in a bath of small, non-chiral, active Brownian particles in two dimensions. 
 %They discussed the role of activeness of the bath particles and the shape of the colloidal inclusions and showed that effective two-body interaction forces between the inclusions vary drastically depending on their shape (and orientation), as also discussed in other recent works (see, e.g., Refs. \cite{Ni:PRL2015,Leite:PRE2016,Smallenburg:2015}).
For disk-shaped inclusions, they showed that the effective two-body force mediated between the inclusions can be strongly repulsive, in stark contrast with the non-active case, where, as noted above, depletion forces are attractive. The repulsive forces were shown to result from  surface accumulation of active particles near the inclusions (a common trait for biological/colloidal active particles at confining boundaries \cite{elgeti2016,wallattraction,wallattraction2,wallattraction3,li2011accumulation,li2011accumulation2,ardekani,Hernandez-Ortiz1,stark-wall,gompper-wall,elgeti2009,elgeti2013,upstream-goldstein,rusconi,elgeti2015run,costanzo,catesupstream,ezhilan,sabass}) and, in particular,  in the intervening region between the inclusions as they are brought to small surface separations  \cite{Harder:2014}. 
%The effective interaction forces vary drastically depending on the shape (and orientation) of the colloidal inclusions as discussed in other recent works as well (see, e.g., Refs. \cite{Ni:PRL2015,Leite:PRE2016,Smallenburg:2015}).

Here, we first revisit the problem of effective two-body interactions between disk-shaped inclusions in an active bath within a commonly used, two-dimensional model of self-propelled particles consistent with the one studied in Ref. \cite{Harder:2014}. Using Brownian Dynamics simulations, and by considering both chiral and non-chiral active baths, we show how the salient, and previously unexplored, aspects of the spatial distribution of active particles, including formation of {\em circular layers} (or `{\em rings}') of active particles around the inclusions, determine the qualitative nature (e.g., repulsive versus attractive) as well as quantitative features (e.g., non-monotonic behavior with distance) of the two-body interaction force profiles. The sequential overlaps of active-particle rings, which in a way create soft, repulsive, `shoulders' \cite{Zihrel1,Zihrel2} around the inclusions, generate the distinct features of the force profiles as the inclusions are brought to small surface separations. Active-particle chirality leads to suppression of particle rings and partial depletion of active particles from the intervening region and the farther proximity of the inclusions. We thus show how particle chirality can engender a crossover in the parameter space between two characteristically different limits of pure active repulsion and pure depletion attraction; hence, linking them as two complementary limits for the system behavior. We also argue that, through their spatial structuring around the inclusions, active particles can also cause significant non-pairwise effects in the interaction forces between the inclusions, for which we provide direct evidence by investigating these effects on the leading three-body level. 

The paper is organized as follows: In Section \ref{sec:model-methods}, we introduce our model and details of our simulations. The distribution of non-chiral and chiral active particles and their influence on the two- and three-body interactions between the inclusions are discussed in Sections \ref{sec:2colloids} and \ref{sec:3colloids}, respectively. The  paper is concluded in  Section \ref{sec:Conclusion}.

%-------------------------------
\begin{figure}[t!]
 \begin{center}
 \includegraphics[width=0.375\textwidth]{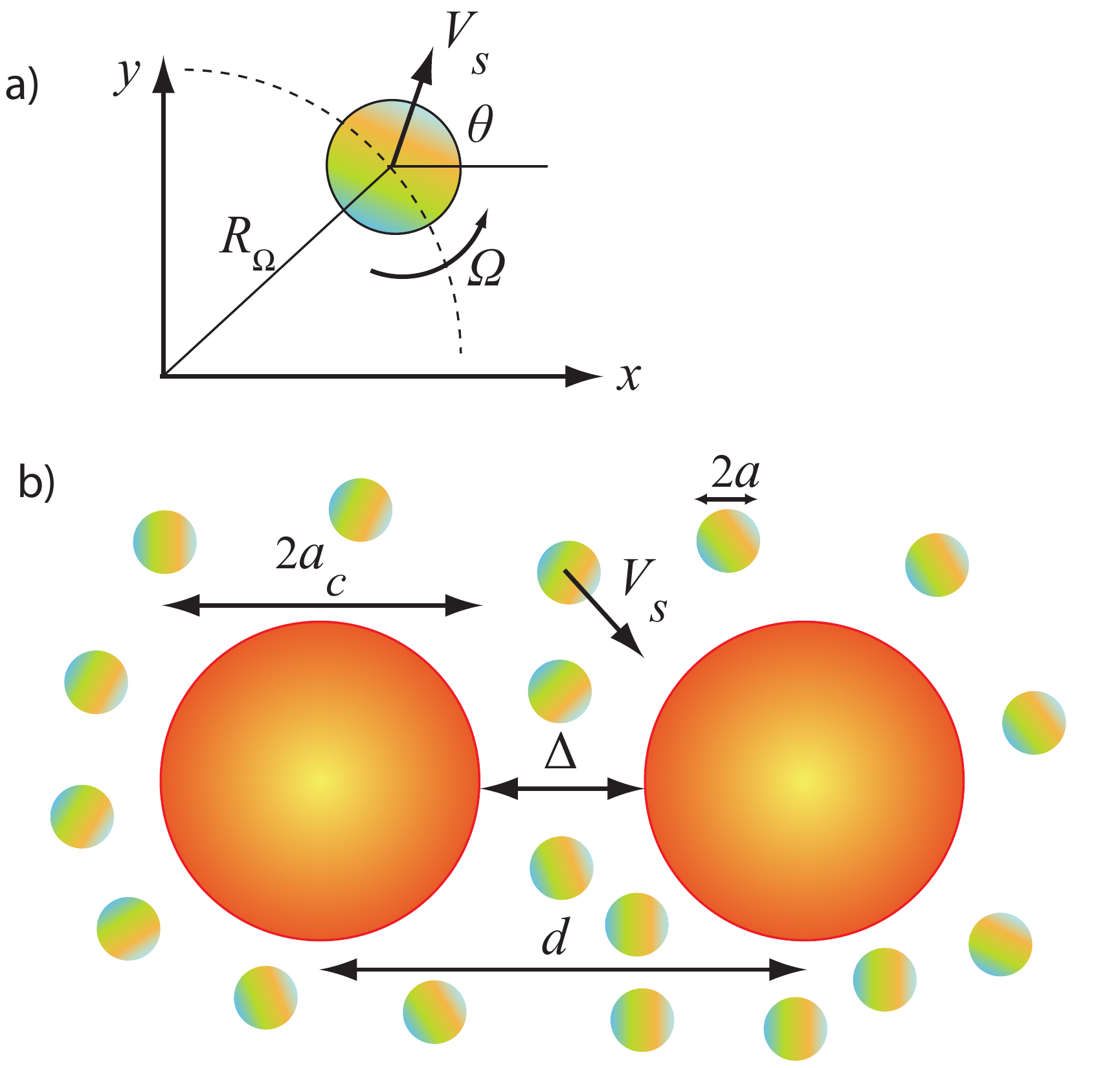}
 \caption{a) Schematic view of a chiral active particle with self-propulsion speed $V_s$ and counterclockwise angular velocity  $\Omega$. b) Two non-active colloidal disks are placed at fixed positions with surface-to-surface distance $\Delta$ in a bath of chiral (or non-chiral) active particles moving according to Eqs. (\ref{Eq:langevin_a}) and (\ref{Eq:langevin_b}).
 }
 \label{fig:schematic}
\end{center}
\end{figure}
%-------------------------------

%SECTION%%%%%%%%%%%%%%%%%%%%%%%%%%%%%%%%%%%%%%%%%%%%%%%%%%%%%%
\section{Model and methods}
\label{sec:model-methods}

%SUBSECTION%%%%%%%%%%%%%%%%%%%%
%\subsection{Model and dynamical equations}
%\label{subsec:model}

Our model consists of two or three identical, non-active and nearly hard colloidal inclusions of radius $a_c$ placed at fixed positions in a bath of identical, self-propelled,  Brownian particles of smaller radius $a<a_c$ and area fraction $\phi$  in two spatial dimensions. Such two-dimensional models have widely been used in the literature as they enable computationally efficient simulations, while capturing key features of active systems  \cite{Lauga:RPP009,Marchetti:RMP2013,Yeomans:EPJ2014,Elgeti:RPP2015,Lauga:ANNREVF2016,golestanian_review,ramaswamyreview,cates_review,Romanczuk:EPJ2012,ZottlStark_review,bechinger_review}.

Active particles are assumed to self-propel at constant speed $V_s$ along their preferred direction of motion ${\mathbf n}=(\cos \theta, \sin \theta)$. The instantaneous configuration of active particles (labelled by $i$) can thus be described by the set of position vectors ${\mathbf r}_i(t) = (x_i(t), y_i(t))$, and orientation angles $\theta_i(t)$ (see Fig. \ref{fig:schematic}). These degrees of freedom are assumed to evolve in time according to the overdamped Langevin equations (see, e.g., Refs. \cite{Marchetti:RMP2013,Elgeti:RPP2015,Romanczuk:EPJ2012,Volpe:2014})
\begin{eqnarray}
\dot{{\mathbf r}}_i &= &V_s{\mathbf n}_i-\mu_T\frac{\partial  U(\{{\mathbf r}_j\})}{\partial {{\mathbf r}_i}}+\sqrt{2 D_T}\, {\boldsymbol \eta}_i(t),
\label{Eq:langevin_a}
\\
\dot{\theta}_i&=&\Omega + \sqrt{2D_R}\, \zeta_i(t),
\label{Eq:langevin_b}
\end{eqnarray}
where $D_T$ and $D_R$ are the (bare) translational and rotational diffusion coefficients of active particles, and $\Omega$ their intrinsic angular velocity, whose sign $S_\Omega=\pm 1$ gives the chirality type, with positive (negative) sign corresponding to counterclockwise (clockwise) angular velocity. Here, ${\boldsymbol \eta}_i(t)$ and $\zeta_i(t)$ are independent, white, Gaussian translational and rotational noises, respectively. They have zero mean,  $ \langle { \eta}_i^\alpha(t)  \rangle=  \langle \zeta_i(t)  \rangle=0$, and two-point time correlations $ \langle {\eta}_i^\alpha(t) {\eta}_j^\beta(t') \rangle=\delta_{ij}\delta_{\alpha\beta}\delta(t-t')$ and  $ \langle \zeta_i(t) \zeta_j(t') \rangle=\delta(t-t')$, with $i, j$ denoting the active particle labels and $\alpha, \beta$ the two Cartesian directions $x, y$. 

We assume the Einstein-Smoluchowski-Sutherland relation between translational mobility and diffusion coefficients  $D_T  = \mu_T k_{\mathrm{B}}T$, where $k_{\mathrm{B}}$ is the Boltzmann constant and $T$ the ambient temperature. This assumption and the choice of thermal (passive) noises are made to ensure that the system approaches its appropriate Boltzmann-weighted equilibrium determined by the potential $U(\{{\mathbf r}_j\})$ in the steady state, when the active self-propulsion is switched off. This enables direct comparisons between non-equilibrium effective interactions generated in an active bath and their equilibrium counterpart when bath particles are non-active. Our model thus excludes athermal noises associated with fluctuations in the self-propulsion and angular velocities  \cite{Romanczuk:EPJ2012,Romanczuk:PRL2011,Grosmann:NJPhys2013,Grosmann:NJPhys2012,Romanczuk:EPJ2015}. 

The potential energy $U$ in Eq. (\ref{Eq:langevin_a}) gives the sum of pair potentials between all particles (including active particles and inclusions) in the system. The pair potential is taken in a Weeks-Chandler-Andersen (WCA) form as
%, or shifted and truncated Lennard-Jones) 
\begin{equation}
 V_{\mathrm{WCA}}({\mathbf r}) =
\left\{\begin{array}{l l}
\!4\epsilon\! \left [ \left ( \frac{\sigma_{\mathrm{eff}}}{|{\mathbf r}|} \right )^{12}\! -2 \left ( \frac{\sigma_{\mathrm{eff}}}{|{\mathbf r}|} \right )^{6}+1\right ] &\,\,\, |{\mathbf r}| \leq \sigma_{\mathrm{eff}}, \\ 
\!0 &\,\,\, |{\mathbf r}| > \sigma_{\mathrm{eff}},
\end{array}\right.
\end{equation}
where $|{\mathbf r}|$ is the center-to-center distance between the considered pair of particles with $\sigma_{\mathrm{eff}}=2a$, when the particles considered are both of active particles,   $\sigma_{\mathrm{eff}}=2a_c$, when the pair are both of inclusions, and $\sigma_{\mathrm{eff}}=a+a_c$, when one of the particles is an active particle and the other one is an inclusion. In all these cases, the interaction energy strength is taken to be the same and equal to $\epsilon$.

%SUBSECTION%%%%%%%%%%%%%%%%%%%%
\subsection{Simulations: Methods and parameters}
\label{subsec:sim}

In order to proceed, we use a dimensionless representation by rescaling the units of length and time as
\begin{equation}
\tilde x = \frac{x}{a},\quad \tilde y = \frac{y}{a},\quad \tilde t = \frac{D_T t}{a^{2}}.
\label{Eq:dimensionless}
\end{equation}

Equations (\ref{Eq:langevin_a}) and (\ref{Eq:langevin_b}) can then be solved numerically using Brownian Dynamics methods by rewriting them in dimensionless and discrete form for time evolution over a sufficiently small time step $\Delta \tilde t$ as 
\begin{eqnarray}
&&\tilde x_i(\tilde t +\Delta \tilde t ) = \tilde x_i(\tilde t)+[Pe_s \cos \theta_i(\tilde t) + \tilde f^x_i(\tilde t)] \Delta \tilde t + \sqrt{2 \Delta \tilde t}\,R^x_i\nonumber\\
\label{Eq:2Ddimensionless-1}
\\
&&\tilde y_i(\tilde t +\Delta \tilde t ) = \tilde y_i(\tilde t)+[Pe_s \sin \theta_i(\tilde t) + \tilde f^y_i(\tilde t)] \Delta \tilde t  + \sqrt{2 \Delta \tilde t}\,R^y_i\nonumber\\
\label{Eq:2Ddimensionless-2}
\\
&&\theta_i (\tilde t +\Delta \tilde t ) = \theta_i(\tilde t) + \chi\, S_\Omega \Gamma \Delta \tilde t + \sqrt{2 \chi \Delta \tilde t}\,R^\theta_i,
\label{Eq:2Ddimensionless-3}
\end{eqnarray}
where $\tilde f^x_i = -\partial \tilde U/\partial \tilde x_i$ and  $\tilde f^y_i = -\partial \tilde U/\partial \tilde y_i$ are the Cartesian components of the dimensionless force derived from the rescaled potential $\tilde U=U/(k_{\mathrm{B}}T)$, and $R^x_i$,  $R^y_i$ and $R^\theta_i$ are independent Gaussian-distributed random numbers with zero mean and unit variance. In Eq. (\ref{Eq:2Ddimensionless-3}), we have defined  $\chi = a^2D_R/D_T$,  which, using the fact that in the low-Reynolds-number (Stokes) regime the translational and rotational diffusion coefficients for no-slip spherical particles satisfy the relation $D_R=3D_T/4a^2$ \cite{happelbook}, can be fixed conventionally as $\chi=3/4$. We have also defined the {\em chirality strength parameter} as the dimensionless magnitude of the (intrinsic) particle angular velocity, 
\begin{equation}
\Gamma = \frac{|\Omega|}{D_R},
\label{eq:Gamma}
\end{equation}
and the so-called {\em swim P\'eclet number} as % \cite{Theurkauff:PRL2012}
\begin{equation}
Pe_s=\frac{a V_s}{D_T} = \frac{3 V_s}{4D_R a}. 
\end{equation}
This latter quantity gives the ratio of the characteristic timescale of the translational (or rotational) diffusion of the active particles, $a^2/D_T$ (or $1/D_R$), to the characteristic timescale of their active self-propulsion (swim), $a/V_s$. Note also that, in the present context, the sign of the angular velocity of active particles is irrelevant and only its magnitude, $|\Omega|$, will play a role. 

In rescaled units, the system is described by the size ratio $a_c/a$, the rescaled area fraction $\phi a^2$, the swim P\'eclet number $Pe_s$, the chirality strength parameter $\Gamma$, and the rescaled center-to-center distance between adjacent inclusions $\tilde d = d/a$ or, equivalently, their rescaled surface-to-surface distance $\tilde \Delta = \Delta/a$ with $\Delta=d-2a_c$ (see Fig. \ref{fig:schematic}). Our focus will be on the roles of self-propulsion and chirality strengths as our main control parameters; we will then fix other parameter values as $\epsilon/(k_{\textup{B}}T)=10$, $a_c/a = 5$ and $\phi a^2=0.1$, consistent with those chosen in  Ref.  \cite{Harder:2014} (the effects due to varying size ratio and area fraction for active particles and inclusions will be discussed in the more general context of active mixtures elsewhere \cite{Zaeifi:2017}). In our simulations, $Pe_s$ is increased from 0 up to around 45 and $\Gamma$ from 0 up to around 19, spanning a wide range of experimentally accessible, actual parameter values; e.g., the cases with $Pe_s = 22.8$ and $\Gamma = 19.1$, which will be considered later, can be mapped to $a = 1\, \mu{\mathrm{m}}$, $a_c = 5\, \mu{\mathrm{m}}$, $V_s \simeq 5\, \mu{\mathrm{m}}\cdot{\mathrm{s}}^{-1}$, $D_T \simeq 0.22\, \mu{\mathrm{m}}^2\cdot{\mathrm{s}}^{-1}$, $D_R \simeq 0.16\,{\mathrm{s}}^{-1}$ and $|\Omega| \simeq \pi\,\,{\mathrm{s}}^{-1}$ in an aqueous medium with shear viscosity $\eta = 0.001\, {\mathrm{Pa\cdot s}}$ \cite{Volpe:2014}. It is also worth mentioning that active particles can exhibit a wide range of chirality strengths in experiments; examples include Janus doublets ($\Gamma\simeq 14 - 27$) \cite{Golestanian:PRE2010}, self-propelled rods ($\Gamma\simeq 8$) \cite{Shelley:2013} and active L-shaped particles ($\Gamma\simeq 200$) \cite{Bechinger:PRL2013}. Therefore, while our focus will primarily be on the generic aspects of the considered model, the parameters are varied within a realistic range of values. 

% mutant {\em E. coli} can have angular velocity of $|\Omega|\simeq 0.3 - 6.0\,{\mathrm{s}}^{-1}$ [86], giving chirality strength parameter in the range with $\Gamma\simeq 1.9 - 37.5$ \cite{wallattraction2}, (using the known bare rotational diffusivity $D_R\simeq 0.16\,{\mathrm{s}}^{-1}$, as noted in the manuscript), Janus doublets with $\Gamma\simeq 14 - 27$ \cite{Golestanian:PRE2010}, self-propelled rods with $\Gamma\simeq 8$ \cite{Shelley:2013}, and L-shaped active particles with $\Gamma\simeq 200$ \cite{Bechinger:PRL2013}. 
% (due to their very small rotational diffusivities) [62]

%a few examples of {\em experimental realization} of chiral swimmers with reference numbers from our revised version of the manuscript:  Self-propelled rods as discussed in Refs. [23,65], Janus doublets [31,68], 
%%Janus particles with asymmetries in their structure \cite{Marine:PRE2013,Wheat:Langmuir2010},
%asymmetric L-shaped microswimmer [30],  mutant %(HCB437) 
%{\em E. coli} [86] and sperm cells [66]. There are other examples of such swimmers, such as those considered in Refs.  \cite{Mirkovic:SMLL2010,Boymelgreen:PRE2014, Marine:PRE2013,Wheat:Langmuir2010}. 

Our simulations typically run for $10^6-10^8$ time steps (with typical time-step size chosen as $\Delta \tilde t\simeq 10^{-4}$) with $10^6$ steps used initially for relaxation purposes and the rest used for computing averaged quantities; the results are further averaged over about 20 statistically independent samples after the system has reached a steady state. For the most part, the simulations are performed using one hundred active bath particles distributed in random initial positions in a square box with periodic boundary conditions, whose lateral size is adjusted according to the given area fraction (test simulations for systems with up to 300-400 particles show only quantitative differences of at most 10\%, while qualitative aspects of our results remain unchanged). In the plots shown later, computed error bars are typically smaller than the size of symbols.

%SUBSECTION%%%%%%%%%%%%%%%%%%%%
\subsection{Forces acting on inclusions}
\label{subsec:force_sim}

The net force acting on a fixed inclusion (in the absence or presence of other fixed inclusions in the bath) follows from the averaged sum of instantaneous forces exerted on it by bath particle collisions. To introduce some of the terminology that we shall use later, we describe here the method of calculating the two-body forces to be analyzed in Section \ref{sec:2colloids}. The method of evaluating three-body forces will be discussed  in Section \ref{sec:3colloids}.

Since the active bath is homogeneous and isotropic, the {\em net} force acting on a single inclusion due to active particle collisions in the bath turns out to be zero on average (within our simulation margin of error). Hence, in the case of two fixed inclusions, the {\em net} force, which is found to be non-zero and to act on each of the inclusions with equal magnitude and in opposite directions along the $x$ (center-to-center) axis, can be interpreted as the effective, two-body, {\em interaction} force mediated between them by the bath. We conventionally denote the rescaled force acting on the inclusion placed on the right in Fig. \ref{fig:schematic} as   $ \tilde {\mathbf F}_2  =  \tilde F_2 \hat{\mathbf x}$; thus, an attractive (repulsive) interaction force is represented by a negative (positive) force amplitude $\tilde F_2$. We calculate this interaction force using $\tilde {\mathbf F}_2 = \sum_i \langle\tilde {\mathbf f}_i\rangle$, where the brackets $\langle\cdots\rangle$ denote the average taken over many simulated configurations and $\tilde {\mathbf f_i}=(\tilde f_i^x, \tilde f_i^y)$ denote instantaneous force components imparted on the mentioned inclusion from the $i$th active particle; these force components follow from the respective  WCA interaction potentials discussed in Section \ref{subsec:sim}.  

%-------------------------------
\begin{figure}[t!]
\begin{center}
\includegraphics[width=0.4\textwidth]{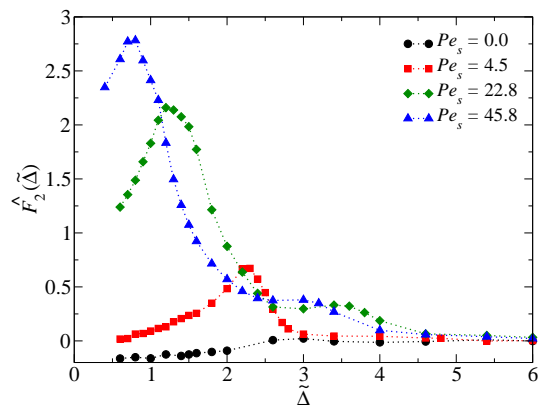}
\caption{Rescaled, effective, two-body force amplitude $\hat F_2$ (as defined in the text), acting on each of the two juxtaposed colloidal inclusions in a non-chiral ($\Gamma=0$) active bath (Fig. \ref{fig:schematic}), is shown as a function of  the rescaled surface-to-surface distance, $\tilde \Delta$, for different values of the P\'eclet number $Pe_s$, as indicated on the graph. The curves are  guides to the eye. 
}
\label{fig:tot_force_2colloids}
\end{center}
\end{figure}
%-------------------------------

%SECTION%%%%%%%%%%%%%%%%%%%%%%%%%%%%%%%%%%%%%%%%%%%%%%%%%%%%%%
\section{Effective two-body interactions}
\label{sec:2colloids}

%SUBSECTION%%%%%%%%%%%%%%%%%%%%
\subsection{Non-chiral active bath}
\label{subsec:2colloids-nonchiral}

We start our analysis of effective two-body interactions by considering a system of two fixed inclusions in a bath of {\em non-chiral} active particles. Figure \ref{fig:tot_force_2colloids} shows the results for the rescaled two-body force amplitude $\tilde F_2$, as defined in Section \ref{subsec:force_sim}, as a function of  the rescaled surface-to-surface distance between the inclusions, $\tilde \Delta$. We show the results for a few different values of the swim P\'eclet number, $Pe_s$, including the non-active case with $Pe_s=0$. For the sake of demonstration, the results are divided by $Pe_s+1$ to display the quantity $\hat F_2 \equiv  \tilde F_2/(Pe_s+1)$.
 
As seen in the figure, the two-body interaction force turns from its typical non-active (equilibrium) form, representing a relatively short-ranged, attractive depletion force (black filled circles) to a much stronger repulsive force with a longer range of action (of a few active-particle radii in surface separations) as $Pe_s$, or the self-propulsion strength, is increased. This is in agreement with the findings in Ref. \cite{Harder:2014}. Our data, however, resolve the behavior of the force in more detail around the first peak and reveals the presence of a secondary hump (which appears more like a plateau region for the give parameter values in the figure) at larger separations, when the swim P\'eclet number is sufficiently large. The first peak occurs at rescaled surface-to-surface distances $\tilde \Delta\lesssim 2$ (or, equivalently, $\Delta\lesssim 2a$) and its location shifts to smaller values as $Pe_s$ is increased. The somewhat peculiar non-monotonic behavior of the two-body force profiles and the presence of a secondary hump, which is not reported in Ref. \cite{Harder:2014}, can be understood by looking more closely at the steady-state distribution of active particles around the inclusions, giving insight into the mechanism underlying the salient features of the force profiles.

%-------------------------------
\begin{figure}[t!]
\begin{center}
\includegraphics[width=0.4\textwidth]{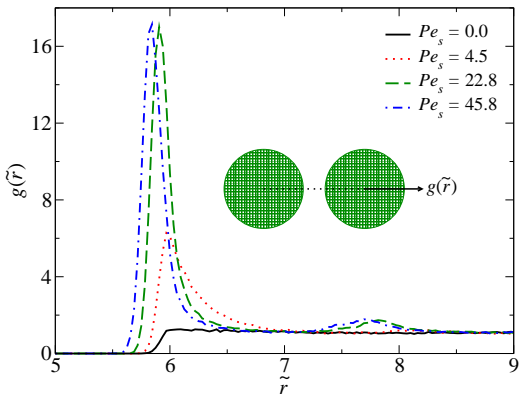}
\caption{Active particle-inclusion pdf $g(\tilde r)$ (see the text for the definition) in the non-chiral two-inclusion system for fixed $\tilde{\Delta} = 0.4$ and different values of $Pe_s$ as shown on the graph. 
} 
\label{fig:gr_2colloids}
\end{center}
\end{figure}
%-------------------------------

We define $g(\tilde r)$ as the probability of finding an active particle at a distance $\tilde r$ from the center of one of the inclusions (e.g., the inclusion on the right in Fig. \ref{fig:gr_2colloids}, inset) along the center-to-center axis in the {\em outward} direction pointing toward the bulk (this particular direction is chosen as the influence of the other inclusion on $g(\tilde r)$ along this direction is negligible). For the sake of brevity, we refer to $g(\tilde r)$ simply as the {\em active particle-inclusion pair distribution function}, or the pdf. As seen in Fig. \ref{fig:gr_2colloids}, this quantity exhibits two peaks, indicating a layered structure for  the spatial distribution of active particles at sufficiently large $Pe_s$. The first peak in $g(\tilde r)$ indicates a high-density layer, or the {\em primary ring}, of active particles at a close distance from the inclusion surface; this distance tends to roughly one active-particle radius as $Pe_s$ is increased, while the second peak in $g(\tilde r)$, indicating a less populated {\em secondary ring} of active particles, is found at a larger distance (specifically, for $Pe_s=45.8$, the primary and secondary rings are found at separations $\tilde r_1 - a_c/a\simeq 0.85$ and $\tilde r_2- a_c/a\simeq 2.8$ from the inclusion surface, respectively). As noted previously, active particles are known to exhibit strong accumulation near surface boundaries \cite{elgeti2016,wallattraction,wallattraction2,wallattraction3,li2011accumulation,li2011accumulation2,ardekani,Hernandez-Ortiz1,stark-wall,gompper-wall,elgeti2009,elgeti2013,upstream-goldstein,rusconi,elgeti2015run,costanzo,catesupstream,ezhilan,sabass}; this tendency is opposed, in the present context, by the steric repulsions between the particles, giving rise to their mentioned layer ordering in the high density regions around the inclusions. 

\begin{figure}[t!]
\begin{center}
\includegraphics[width=0.475\textwidth]{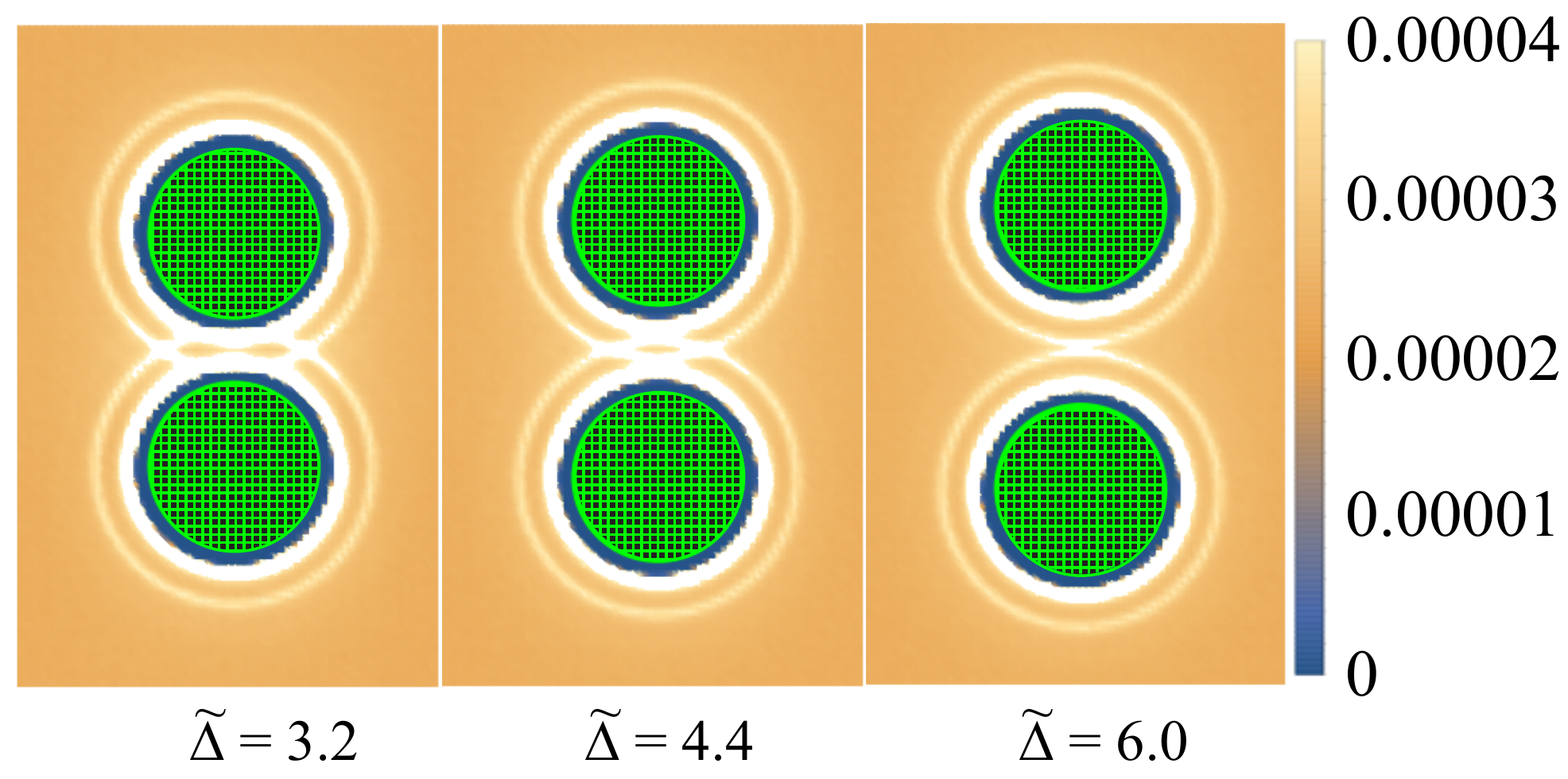}
\caption{Steady-state density maps of non-chiral active particles around colloidal inclusions are shown for three different values of $\tilde \Delta$ 
%$\tilde \Delta = 3.2$, 4.4 and 6.0 (from left to right) 
and $Pe_s = 45.8$.  The green disks show the areas occupied by the two inclusions and the thin (dark blue) circular areas attached to the inclusion disks are regions from which active particles are further excluded due to their finite radius. The active-particle-populated layers appear as white (primary) and yellow (secondary) rings, respectively.  
} 
\label{fig:dens_2colloids}
\end{center}
\end{figure}

The angular dependence of the simulated active particle distribution is shown in Fig. \ref{fig:dens_2colloids}, which complements the information obtained from $g(\tilde r)$ by presenting the two-dimensional view of the active-particle density profiles around and in between the two inclusions. In this figure, the circular areas excluded by the inclusions themselves are shown as green disks. The thin (dark blue) circular areas (of thickness equal to one active-particle radius) that are attached to the inclusion disks are regions from which the active particles are further excluded due to their finite radius. The white areas with high active-particle density correspond to the primary rings around the inclusions and also the regions of overlap between the secondary rings, appearing themselves in yellow. It is worth mentioning here that each inclusion with its two active-particle rings can be thought of as a single colloidal unit with a {\em hard core} and a {\em soft repulsive shoulder}. Qualitatively similar scenarios, incorporating hard-core/soft-shoulder pair potentials, have been considered in the equilibrium context of non-active colloids and shown to result in a diverse phase diagram \cite{Zihrel1,Zihrel2}. Those findings may, however, be inapplicable to the present case, in which the circular soft-ring zones due to active-particle layering form only in a non-equilibrium steady state.

%-------------------------------
%-------------------------------

%-------------------------------
\begin{figure}[t!]
\begin{center}
\includegraphics[width=0.4\textwidth]{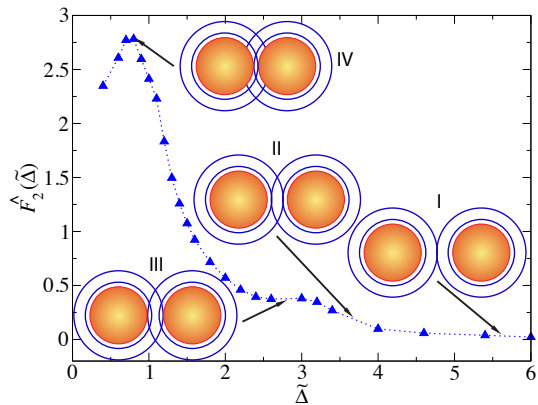}
\caption{Schematic representation of two apposed inclusions and their active-particle rings shown along with the two-body force profile for the exemplary case of $Pe_s = 45.8$ (data are reproduced from Fig. \ref{fig:tot_force_2colloids}). Arrows point to the values of the rescaled surface-to-surface separation of the inclusions, $\tilde \Delta$, corresponding to the four schematics pictures. The magnitude of the effective force acting on the inclusions increases as $\tilde \Delta$ is decreased from I to IV, due to various levels of overlap between the primary and secondary rings of the inclusions.  
} 
\label{fig:schematic_rings}
\end{center}
\end{figure}
%-------------------------------

The non-monotonic behavior of the two-body interaction force profiles, and their increased range of action at elevated $Pe_s$ (Fig. \ref{fig:tot_force_2colloids}), can be understood based on sequential overlaps (or intersections) occurring between the active-particle rings as the two inclusions are brought together. This is shown schematically in Fig. \ref{fig:schematic_rings}, where we re-plot the interaction force profile in the exemplary case of $Pe_s = 45.8$ together with schematic illustrations of two apposed complexes comprising a central inclusion and its two active-particle rings at four indicative surface-to-surface distances. At large separations, the two complexes are decoupled and they begin to interact with a non-vanishing effective force only when their secondary  rings come into contact; this corresponds to configuration I as indicated on the graph (occurring at $\tilde \Delta\simeq 5.6$). As $\tilde \Delta$ is further decreased, the effective repulsion between the inclusions is expected to increase as the secondary ring associated with one inclusions intersects the primary ring of another inclusion (configuration II; $\tilde \Delta\simeq 3.65$). This results from an increased build-up of active particle density at regions of ring intersection between the inclusions (see also the left panel in Fig. \ref{fig:dens_2colloids}) and, as a result, increased collision forces from active particles pushing the inclusions apart. After this point, one can expect only a weak change in the magnitude of the repulsion (leading to the secondary hump or the plateau-like region in the force profile) at about the separation, where the secondary ring of one inclusion comes into contact with the surface of another inclusion (configuration III; $\tilde \Delta\simeq 2.8$). The repulsive force is expected to sharply increase as $\tilde \Delta$ is squeezed down to smaller separations, where the primary ring of one inclusion also comes into contact with the surface of another inclusion (configuration IV; $\tilde \Delta\simeq 0.85$), creating  even larger collision forces from active particles pushing the inclusions apart. This will in fact be the global maximum in the effective force profile since active particles are excluded from the narrow intervening region between the inclusions as $\tilde \Delta$ is decreased even further. 

%SUBSECTION%%%%%%%%%%%%%%%%%%%%
\subsection{Chiral active bath}
\label{subsec:2colloids-chiral}

We now turn to the case of {\em chiral} active particles  in a system with {\em two} fixed colloidal inclusions. Chirality of the active particles adds a new characteristic length scale to the problem $R_{\Omega} = V_s/|\Omega|$ (see Fig. \ref{fig:schematic}), representing the characteristic radius of a typical  circular arc traversed by the particles (note that, in the present context, active particles do not move on perfect circular trajectories due to collisions with other particles and the rotational Brownian noise). Chirality effects are expected to dominate when $R_{\Omega}$ becomes smaller than the swim (run) length $\ell_{\textrm{run}}=V_s/D_R$  \cite{Xue:EPL2015,Xue:EPJST2014}, or equivalently, when the timescale associate with active-particle chirality $|\Omega|^{-1}$ is smaller than that of its rotational diffusion $1/D_R$, as, otherwise, the chirality effects are masked by the rotational diffusion process. These conditions translate to having a large chirality strength parameter $\Gamma\gg 1$. To gain a more systematic understanding of the chirality effects in the present context, we run Brownian Dynamics simulations similar to those reported in Section \ref{subsec:2colloids-nonchiral} but  assuming here that $\Gamma$ is finite (ranging from 0 up to around 19, corresponding to angular velocity magnitudes $|\Omega|$ from 0 up to around $\pi \,{\mathrm{s}}^{-1}$, when $D_R \simeq 0.16\,{\mathrm{s}}^{-1}$ \cite{Volpe:2014}).  

%Presence of torque forces particles to move in a circular motion in which the radius of the circle will be dependent on the magnitude of torque and the value of self-propulsion ($R_\Omega = V_s/\left|\omega\right|$). As the applied torque gets bigger values the radius of the circle in the motion will be decreased and as the self-propulsion increases the radius of arc will increase. 

%-------------------------------
\begin{figure}[t!]
 \begin{center}
\includegraphics[width=0.4\textwidth]{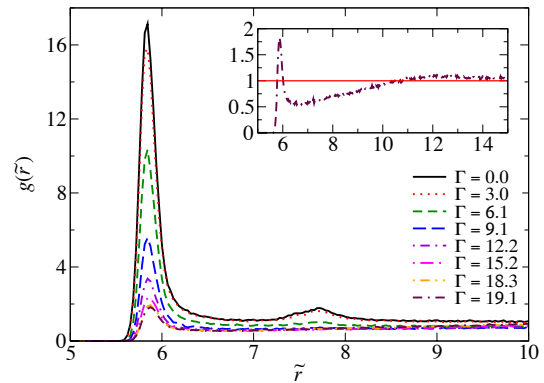}
\caption{Same as Fig. \ref{fig:gr_2colloids} but plotted here is the active particle-inclusion pdf $g(\tilde r)$ (see Section \ref{subsec:2colloids-nonchiral} for the definition) for fixed $\tilde{\Delta} = 1$, $Pe_s=45.8$ and different values of the chirality strength parameter $\Gamma$,  as shown on the graph. 
% $\Gamma= 0, 3.0, 6.1, 9.1, 12.2, 15.2, 18.3$ and 19.1 (corresponding to $|\Omega|\simeq 0, 0.5, 1.0, 1.5, 2.0, 2.5, 3.0, \pi\,{\mathrm{s}}^{-1}$, respectively, when one assumes  $D_R \simeq 0.16\,{\mathrm{s}}^{-1}$). 
}
\label{fig:gr_2colloids_chiral}
\end{center}
\end{figure}
%-------------------------------

%-------------------------------
\begin{figure}[t!]
 \begin{center}
\includegraphics[width=0.275\textwidth]{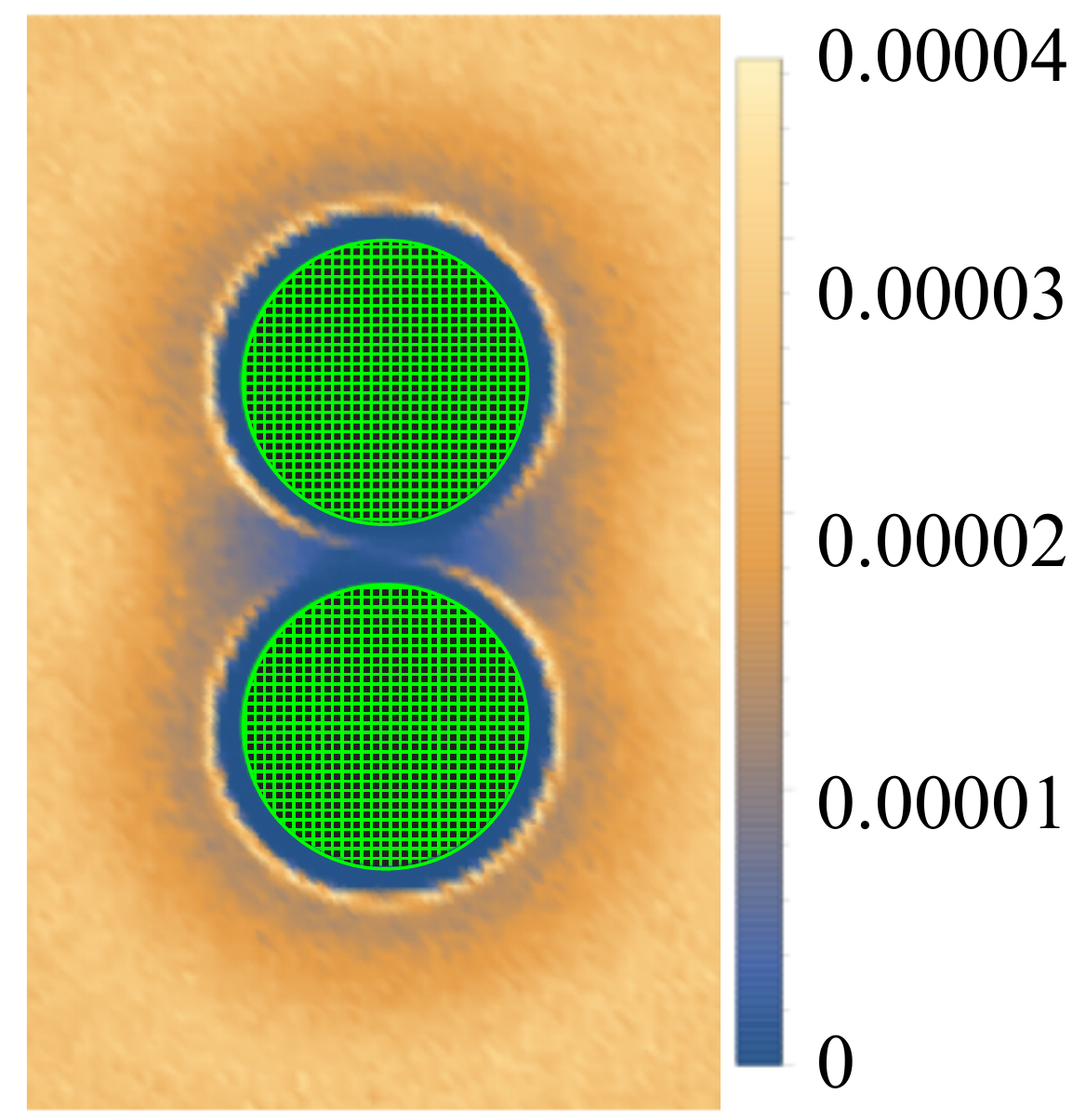}
  \caption{
 Same as Fig. \ref{fig:dens_2colloids} but plotted  is density map of chiral active particles around two inclusions for $\tilde{\Delta} = 3.2 $, $Pe_s = 45.8$ and $\Gamma = 19.1$. See the text and Fig. \ref{fig:dens_2colloids} for more details. 
}
  \label{fig:dens_2colloids_chiral}
  \end{center}
\end{figure}
%-------------------------------

As seen in Fig. \ref{fig:gr_2colloids_chiral}, the peaks in the active particle-inclusion pdf are largely suppressed upon increasing the chirality strength parameter.    Interestingly, however, we find an extended range of separations (Fig. \ref{fig:gr_2colloids_chiral}, inset) over which the pdf drops below its bulk value, representing a region of partial depletion right after the primary ring (the width of this depletion region is of the order of the rescaled radius of curvature, i.e., $R_{\Omega}/a = 4Pe_s/(3\Gamma)\simeq 3.2$ for the parameter values in the inset). The density map of Fig. \ref{fig:dens_2colloids_chiral} gives a clearer view of the primary (yellow) rings at the vicinity of the inclusions and the relatively thick, circular, partial depletion zones (in darker hazy colors) at their farther proximity; the overlap between these latter zones gives a narrow intervening region of strong particle depletion (in dark blue) between the inclusions, where their primary rings disappear, too. 
% (note that, at these parameter values, a small but discernible fraction of the active particles still accumulate next to the inclusion surfaces as demonstrated by yellow rings in the figure, or by the small short-distant peak in the active particle-inclusion pdf in Fig \ref{fig:gr_2colloids_chiral}). 
% (compare also Fig. \ref{fig:dens_2colloids_chiral} and the left panel in Fig. \ref{fig:dens_2colloids}). 

%-------------------------------
\begin{figure}[t!]
 \begin{center}
\includegraphics[width=0.4\textwidth]{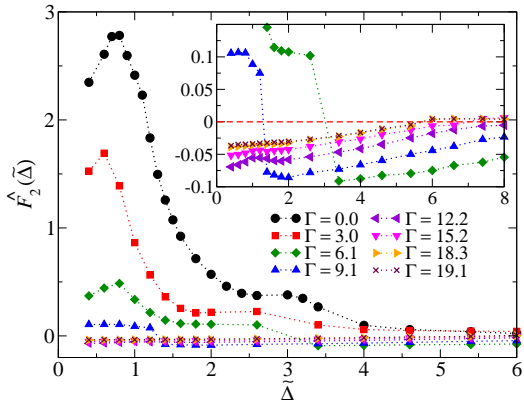}
\caption{Same as Fig. \ref{fig:tot_force_2colloids} but plotted here is the rescaled, effective, two-body force amplitude $\hat F_2$ (see Section \ref{subsec:2colloids-nonchiral} for the definition) as a function of $\tilde \Delta$ for $Pe_s=45.8$ and different values of the chirality strength parameter $\Gamma$, as shown on the graph. The inset shows a closer view of the regime, where the force becomes negative (attractive) for large enough $\Gamma$. The lines connecting the symbols are plotted as guides to the eye.
}
\label{fig:tot_force_2colloids_chiral}
\end{center}
\end{figure}
%-------------------------------

As a consequence of the changes in the spatial distribution of active particles, the effective two-body force between the inclusions varies drastically as $\Gamma$ is increased; see Fig. \ref{fig:tot_force_2colloids_chiral}. The repulsive small-distance peak in the effective force is suppressed and so is the force magnitude over the plateau-like region at larger surface-to-surface distances $\tilde \Delta$. The plateau-like region is also moved to smaller separations and, at sufficiently large chirality strength (data for $\Gamma \geq 6.1$ in the graph), it is followed by a relatively sharp drop to a larger-distance region, where the interaction force between the inclusions becomes attractive (negative). A closer view of this behavior is shown in the inset of Fig. \ref{fig:tot_force_2colloids_chiral}. 

The effective attraction found at sufficiently large $\Gamma$ thus appears to display some of the qualitative features of non-active (equilibrium) depletion attractions \cite{Lekkerkerker:2011}. In fact, one can show using systematic arguments (albeit in a different context \cite{Tayeb-unpub}) that, when the limit $\Gamma\rightarrow \infty$ is taken, the steady-state properties of a system of self-propelled Brownian particles reduce to their corresponding equilibrium values (as obtained by setting the self-propulsion strength equal to zero). %and assuming that such an equilibrium state exits 
It should, however,  be noted that, within our model, chirality effects will be seen only when the particles are active. Also, the resulting effective attraction at finite, even though large chirality strength, still shows significant deviations from its non-active limiting form; these deviations can be traced back to the structure of particle depletion zones, which remain different from their limiting non-active form (thin, dark blue,  circular areas attached to the inclusions Fig. \ref{fig:dens_2colloids_chiral}). Hence, in addition to the fact that the {\em active, chirality-induced, depletion attraction} reported here represents a qualitatively distinct state of non-equilibrium, it also exhibits a range of action a few times larger than its characteristically short-ranged, equilibrium  counterpart \cite{Lekkerkerker:2011} (compare data in Fig. \ref{fig:tot_force_2colloids_chiral}, inset, with $Pe_s=0$ in Fig. \ref{fig:tot_force_2colloids}). 

Our data also shows a non-monotonic trend in the behavior of the attractive force as a function of $\Gamma$ (Fig. \ref{fig:tot_force_2colloids_chiral},  inset): Increasing $\Gamma$, after the force mediated between the inclusions turns attractive, slightly decreases the force magnitude and, as noted above, gradually shifts the attractive region to smaller $\tilde \Delta$ as $\Gamma$ is further increased. 
%By increasing the value of torque in a constant self-propulsion, causes an increase in the attractive force between colloidal disk. However, if the value of torque increased too much as a result the radius of circular arc decreases and tend to suppress the chirality effect. To be able to see the effect of chirality at high torque values, it is required to increase the self-propulsion.
%In the strong chirality regime, as noted before, the characteristic length scale $R_{\Omega}$ becomes much shorter than the self-propulsion (run) length and, as a result, the effects due to self-propulsion (swim) are known to become substantially weaker \cite{Xue:EPL2015,Xue:EPJST2014}, in agreement with the general findings of our simulations. 

%-------------------------------
\begin{figure}[t!]
\begin{center}
\includegraphics[width=0.375\textwidth]{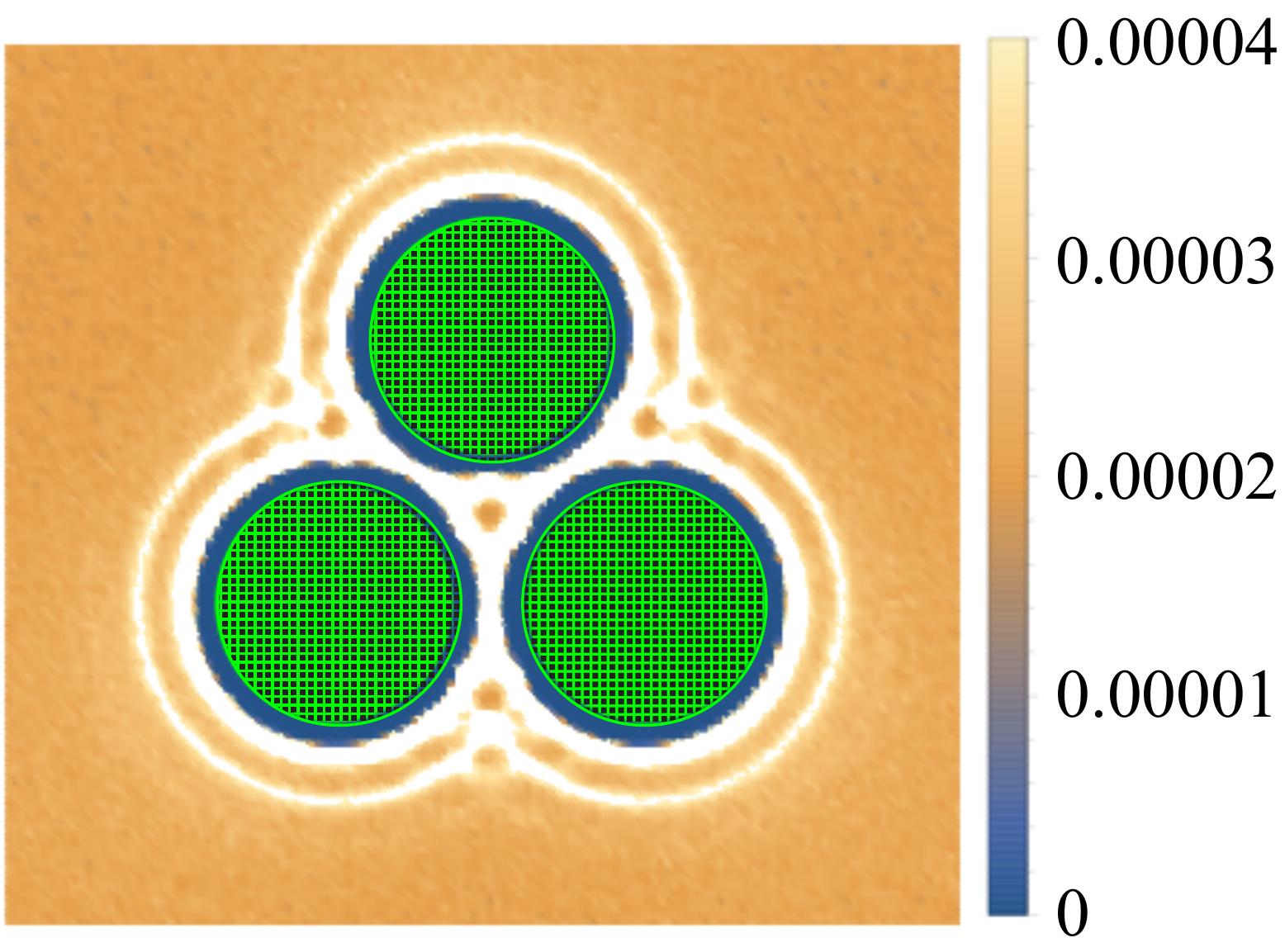} %Dennsity_map_NC_3_Pe_45_omega_0_dist_15}
  \caption{Steady-state density map of non-chiral active particles with $Pe_s = 45.8$  around three colloidal inclusions placed at equal surface-to-surface distance $\tilde \Delta=1.5$ in an equilateral-triangle formation. See Fig. \ref{fig:dens_2colloids} for more details.
  }
  \label{fig:dens_3colloids}
\end{center}
\end{figure}
%-------------------------------

%SECTION%%%%%%%%%%%%%%%%%%%%%%%%%%%%%%%%%%%%%%%%%%%%%%%%%%%%%%
\section{Three-body interactions}
\label{sec:3colloids}

%SUBSECTION%%%%%%%%%%%%%%%%%%%%
\subsection{Non-chiral active bath}
\label{subsec:3colloids-nonchiral}

In non-active systems, depletion interactions between colloidal inclusions in a bath of (smaller) particles are generally known to be non-pairwise, especially, at relatively high volume fractions, where such (typically short-ranged) interactions become important \cite{Lekkerkerker:2011}. Another important factor is the  thickness of the depletion layer around the inclusions: As illustrated in Ref. \cite{Lekkerkerker:2011} (see Fig. 3.8 therein), non-pairwise effects become important when  depletion layers are thick enough (relative to the inclusion size) to exhibit multiple overlap regions as more than two inclusions are brought to small surface separations from one another; in this case, the effective (net) interaction forces acting on the inclusions will involve significant multi-body contributions. The lowest order of multiple overlaps occurs in the case of three inclusions placed at equal surface-to-surface distances creating an equilateral-triangle formation \cite{Lekkerkerker:2011}; this is indeed the case that we shall consider in what follows by first assuming that the active particles are non-chiral ($\Gamma=0$). 

The two-dimensional active-particle density profile around and in between the inclusions in this case is shown in Fig. \ref{fig:dens_3colloids} by taking the surface-to-surface distance of the inclusions as $\tilde \Delta=1.5$ and the P\'eclet number of the active particles as $Pe_s = 45.8$. 
%As seen in Fig. \ref{fig:dens_3colloids}, the basic features of the spatial distribution of (non-chiral) swimmers in the case of three colloidal inclusions in the bath. 
The interesting aspect of the problem in this case is that the relatively thick shoulders comprising two rings of active particles enable large multiple-overlap areas over a range of surface separations comparable to a few active-particle radii. Hence, in analogy with the multiple-overlap mechanism described above \cite{Lekkerkerker:2011}, we may as well anticipate significant non-pairwise three-body contributions to occur due to such overlaps in the present context. 
 
\begin{figure}[t!]
\begin{center}
\includegraphics[width=0.4\textwidth]{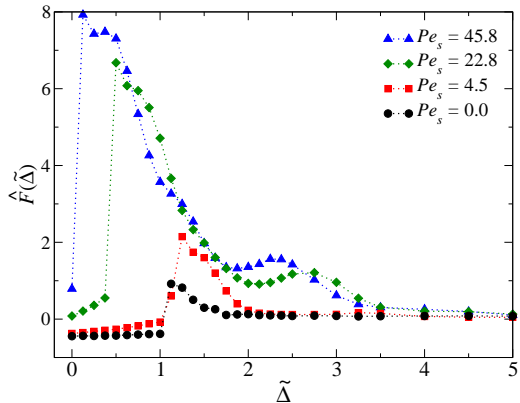} %Dennsity_map_NC_3_Pe_45_omega_0_dist_15}
  \caption{Rescaled, effective, net force $\hat F$, acting on individual inclusions in the three-inclusion configuration (Fig. \ref{fig:dens_3colloids})  in a non-chiral ($\Gamma=0$) active bath, as a function of  the rescaled surface-to-surface distance, $\tilde \Delta$, between adjacent inclusions for different values of $Pe_s$, as indicated on the graph. The dotted lines connecting the symbols are plotted as guides to the eye. 
  }
  \label{fig:tot_force_3colloids}
 \end{center}
\end{figure}

The effective net force acting on individual inclusions in a triangular configuration similar to the one shown in Fig. \ref{fig:dens_3colloids} can be evaluated as a function of their (equal) surface-to-surface distance $\tilde \Delta$, and the swim P\'eclet number $Pe_s$. The net forces experienced by the inclusions are equal in size and are aligned with the bisectors of the angles formed by the triangular configuration of the inclusion centers. The results, in rescaled units and divided also by $Pe_s+1$ (see Section \ref{subsec:2colloids-nonchiral}), are shown in Fig. \ref{fig:tot_force_3colloids}, where negative (positive) values represent forces pointing toward (away from) the center of the triangle, respectively. Comparing this figure with Fig.  \ref{fig:tot_force_2colloids}, the net force acting on a single inclusion is found to show roughly  similar qualitative features  in both the two-inclusion and the three-inclusion configurations. There are, however, remarkable differences between the results in the two cases at a given $Pe_s$: The magnitude of the net force  in the three-inclusion configuration shows an overall increase by a factor of more than two; the secondary hump (which appeared more like a plateau region in the force profiles of the two-inclusion system) is more pronounced in the three-inclusion setting; finally, the net force in the non-active case ($Pe_s=0$) displays a repulsive hump in the three-inclusion system, one that is absent in the two-inclusion case (compare Figs. \ref{fig:tot_force_3colloids} and \ref{fig:tot_force_2colloids}). These differences all point to the presence of sizable three-body effects in the system as we shall discuss below.

%-------------------------------
%-------------------------------

%-------------------------------
\begin{figure}[t!]
\begin{center}
\includegraphics[width=0.4\textwidth]{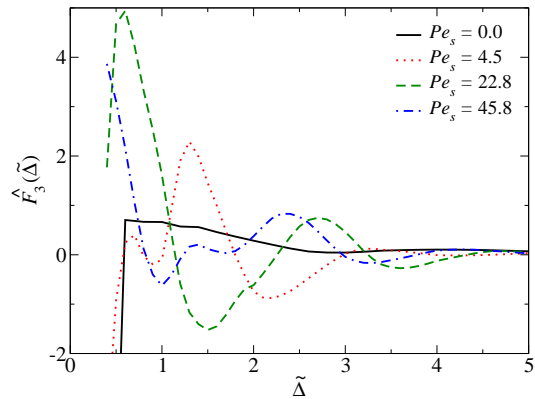} %Dennsity_map_NC_3_Pe_45_omega_0_dist_15}
\caption{Rescaled three-body force $\hat F_3$  (see the text for the definition) that contributes to the net force $\hat F$, acting on individual inclusions in the three-inclusion configuration in a non-chiral active bath (see Fig. \ref{fig:tot_force_3colloids}), is plotted as a function of  $\tilde \Delta$ for different values of $Pe_s$, as shown on the graph. 
  }
  \label{fig:tot_force_3colloids_3body}
 \end{center}
\end{figure}
%-------------------------------

If we denote the effective net force acting on a given inclusion by $F$ and the two-body force acting on this reference inclusion by either of its two neighboring inclusions by $F_2$, then the three-body force follows as 
\begin{equation}
F_3\equiv F - 2F_2\cos \theta,
\label{eq:F_3}
\end{equation} 
where, in the equilateral-triangle arrangement considered here $\theta = \pi/6$. We have also used the fact that, by construction, the two-body force $F_2$ acts along the line connecting the centers of the inclusions in question and that its values can be taken from our results in Fig. \ref{fig:tot_force_2colloids}.  

The resulting three-body force contribution is shown in rescaled units (divided also by $Pe_s+1$) in Fig. \ref{fig:tot_force_3colloids_3body}. (Note that, in our simulations, the set of values for the variable $\tilde \Delta$ at which the net force $\hat F$, Fig. \ref{fig:tot_force_3colloids}, and the two-body force $\hat F_2$, Fig. \ref{fig:tot_force_2colloids}, are calculated, do not necessarily coincide and, thus, $\hat F_3$ is obtained by cubic-spline interpolation of the corresponding data sets in those two figures; hence, reported in Fig. \ref{fig:tot_force_3colloids_3body} are continuous and accurately representing curves rather than individual symbols.)

The simulated three-body force exhibits rapid, sizable variations in sign and magnitude as a function of $\tilde \Delta$ and $Pe_s$, in stark contrast with the behavior found in the case of the two-body interaction force in Fig. \ref{fig:tot_force_2colloids}. The  three-body force becomes strongly repulsive at small separations $\tilde \Delta\lesssim 1$ with magnitudes exceeding those of the two-body force; e.g., in this regime, we find the force ratio $|\hat F_3|/|\hat F_2|\lesssim 2$ for $Pe_s=22.8$ and $|\hat F_3|/|\hat F_2|\lesssim 1.3$ for $Pe_s=45.8$. Our data clearly indicates a non-monotonic behavior for the dependence of $|\hat F_3|$ on  $Pe_s$ at small separations between the inclusions. 

The three-body force profiles as a function of the surface-to-surface distance between the inclusions (Fig. \ref{fig:tot_force_3colloids_3body}) also appear to show two major maxima, whose locations appear to be roughly consistent with the locations of the primary and secondary active-particle rings around the inclusions  (for both $Pe_s=22.8$ and 45.8, these two maxima are found in the intervals $\tilde \Delta\lesssim 1$ and  $2\lesssim\tilde \Delta\lesssim 3$). This correspondence is, however, not precise and the relation between the variations in sign and magnitude of the three-body force and the ring-structure around the inclusions remains to be understood. 

Finally, our data for {\em non-active} bath particles ($Pe_s=0$; Fig. \ref{fig:tot_force_3colloids_3body}) indicate that the repulsive hump, mentioned in our discussion of the net force in Fig. \ref{fig:tot_force_3colloids}, is entirely due to three-body effects. In the two-inclusion system, as the surface separation of inclusions is decreased down to the particle diameter ($\tilde \Delta\simeq 2$), particles are freely depleted from the narrow inter-surface gap between the inclusions, leading to an attractive depletion force without encountering any possible repulsive barrier on the way (Fig. \ref{fig:tot_force_2colloids}). This is not the case for three inclusions being brought to small surface separations, where the wedge-shaped, central void formed in the intervening region between the inclusions can accommodate a finite number of bath particles;  these entrapped particles produce a counteracting outward force barrier, before it is overcome and the entrapped particles are squeezed out upon further packing of the three inclusions together. 

%-------------------------------
\begin{figure}[t!]
\begin{center}
\includegraphics[width=0.375\textwidth]{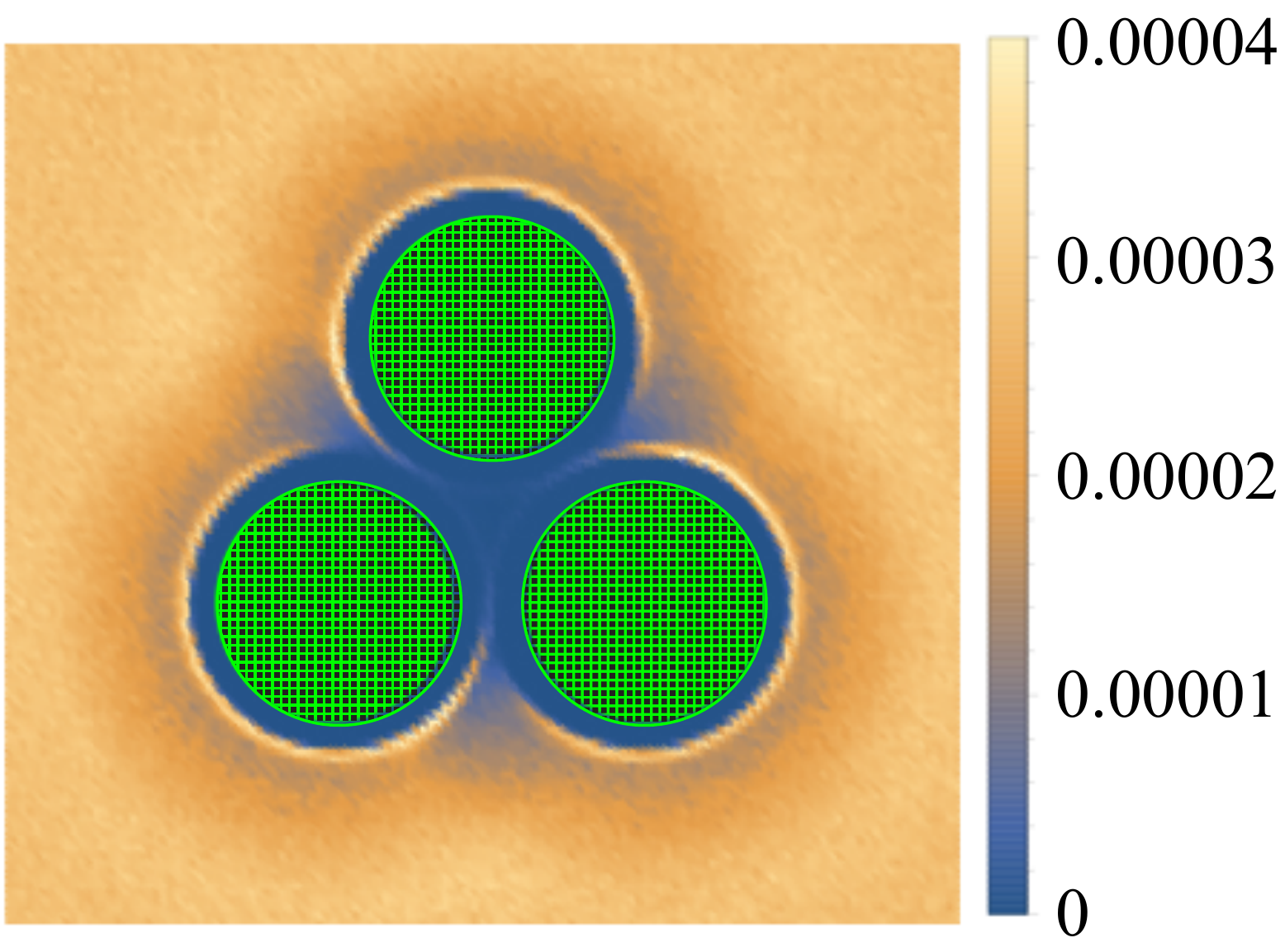}
  \caption{Steady-state density map of chiral active particles  with $Pe_s = 45.8$  and $\Gamma = 19.1$ around three colloidal inclusions at surface-to-surface distances equal to $\tilde \Delta=1.5$ in an equilateral-triangle formation. See Fig. \ref{fig:dens_3colloids}  for more details.
  }
  \label{fig:dens_3colloids_chiral}
 \end{center}
\end{figure}
%-------------------------------

%SUBSECTION%%%%%%%%%%%%%%%%%%%%
\subsection{Chiral active bath}
\label{subsec:3colloids-chiral}

\begin{figure}[t!]
\begin{center}
  \includegraphics[width=0.4\textwidth]{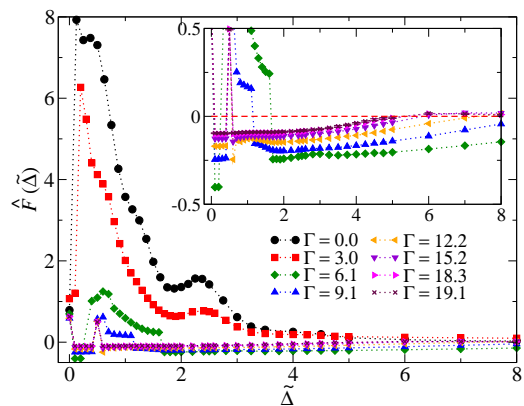}
  \caption{Same as Fig. \ref{fig:tot_force_3colloids} but plotted here is the rescaled, effective, net force $\hat F$ as a function of $\tilde \Delta$ for  $Pe_s=45.8$ and different values of $\Gamma$, as shown on the graph. The inset shows a closer view of the regime,  where the force becomes attractive for large enough $\Gamma$. The dotted lines are guides to the eye.  
  }
  \label{fig:tot_force_3colloids_chiral}
\end{center}
\end{figure}

We proceed by considering the three-body interactions in the case of chiral active particles. The density map of active particles in the three-inclusion system for relatively high values of the swim P\'eclet number and the chirality strength parameter ($Pe_s = 45.8$ and $\Gamma = 19.1$) again shows a primary (yellow) ring of active particles in close vicinity and thick, circular, zones (in darker hazy colors) of partial depletion at the farther proximity of the inclusions, in addition to a central region (in dark blue) of strong depletion. Hence, the effective net force acting on individual inclusions is expected to weaken in strength as $\Gamma$ is increased as confirmed by our data in Fig. \ref{fig:tot_force_3colloids_chiral}. 

Comparing the net force acting on a single inclusion in the two- and three-inclusion configurations (Figs. \ref{fig:tot_force_2colloids_chiral} and \ref{fig:tot_force_3colloids_chiral}), we reach similar conclusions as we did in the non-chiral case in Section \ref{subsec:3colloids-nonchiral}; that is, the magnitude of the net force in the three-inclusion configuration shows an overall increase by a factor of more than two and the secondary hump in the force profile becomes more pronounced, pointing again to significant non-pairwise three-body effects in the system.

%-------------------------------
%-------------------------------

%-------------------------------
\begin{figure}[t!]
\begin{center}
\includegraphics[width=0.4\textwidth]{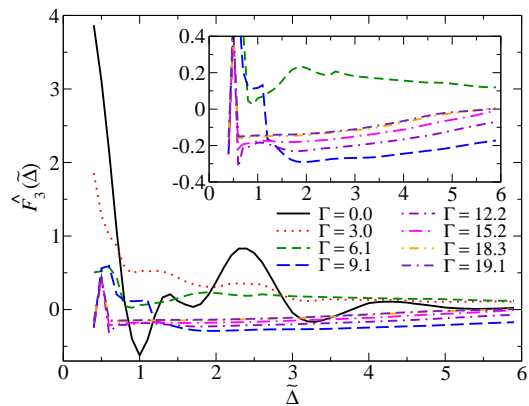} %Dennsity_map_NC_3_Pe_45_omega_0_dist_15}
  \caption{Same as Fig. \ref{fig:tot_force_3colloids_3body} but plotted  is the rescaled three-body force $\hat F_3$ as a function of $\tilde \Delta$ for  $Pe_s=45.8$ and different values of $\Gamma$, as shown on the graph. The inset shows a closer view of the long-range attraction regime for large enough $\Gamma$.
   }
  \label{fig:tot_force_3colloids_3body_chiral}
 \end{center}
\end{figure}
%-------------------------------

The three-body force, which contributes to the effective net force (Fig. \ref{fig:tot_force_3colloids_chiral}) acting on individual inclusions in a bath of chiral active particles, can be calculated using Eq. (\ref{eq:F_3}). The results, in rescaled units and divided also by $Pe_s+1$, are shown  in Fig. \ref{fig:tot_force_3colloids_3body_chiral} for  $Pe_s = 45.8$ and different values of $\Gamma$. Note that the black solid curve in this figure coincides with the blue dashed curve in Fig. \ref{fig:tot_force_3colloids_3body}, representing the non-chiral case ($\Gamma=0$) with its characteristic variations in sign and magnitude as discussed in the previous section. As the chirality strength parameter is increased, these features are suppressed and the three-body force profiles become increasingly more smooth. Thus, particle chirality consistently also weakens the non-pairwise three-body contribution to the effective net force in the three-inclusion configuration. The three-body force becomes attractive over an extended of range surface separations (up to several active-particle radii; see Fig. \ref{fig:tot_force_3colloids_3body_chiral}, inset) for $\Gamma \geq 6.1$, where it can, therefore, be considered as an active  {\em three-body depletion attraction} induced by the particle chirality. 

%The effective three-body attraction at sufficiently large $\Gamma$ (Fig. \ref{fig:tot_force_3colloids_3body_chiral}, inset) exhibits qualitative features similar to the non-active depletion attraction \cite{Lekkerkerker:2011} as, for instance, both are accompanied by a large drop in the density of swimmers in the intervening region between the inclusions as compared to their bulk density (see, e.g., Fig. \ref{fig:dens_3colloids_chiral}). There are, however, important differences between the two cases that were discussed in the two-inclusion case in Section \ref{subsec:2colloids-chiral}. One can, nevertheless, refer to the three-body attraction in this case as active, chirality-induced, {\em three-body} depletion attraction.  

%By increasing the value of torque in a constant self-propulsion, causes an increase in the attractive force between colloidal disk. However, if the value of torque increased too much as a result the radius of circular arc decreases and tend to suppress the chirality effect. To be able to see the effect of chirality at high torque values, it is required to increase the self-propulsion.

%SECTION%%%%%%%%%%%%%%%%%%%%%%%%%%%%%%%%%%%%%%%%%%%%%%%%%%%%%%
\section{Conclusion and Discussion}
\label{sec:Conclusion}

In this paper, we use Brownian Dynamics simulations to study effective interactions generated by an active bath of non-chiral or chiral active particles between large, non-active, colloidal inclusions in the bath. The main emphasis of our study is on the role of the spatial distributions of active particles around the inclusions, and the way they influence the  qualitative nature (such as repulsive versus attractive) as well as the quantitative features (such as non-monotonic behavior with distance) of the two- and three-body interactions that the active bath particles mediate between the inclusions. %We use Brownian Dynamics simulations within a standard two-dimensional model of active particles taken as self-propelled, nearly hard, Brownian disks. 

In the case of a {\em non-chiral} active bath, the effective two-body force profiles are found to exhibit a non-monotonic behavior with finer details than previously reported \cite{Harder:2014}. We provide new insight into the behavior of the force profiles by establishing the formation, and the sequential overlaps, of rings of active particles around the inclusions as the root cause for the salient features of the force profiles that include a primary maximum and a secondary hump (or plateau-like region). The active-particle rings, on the other hand,  create relatively thick, soft shoulders around the hard-core inclusions, enabling them to generate multiple-overlap regions as the inclusions are brought together; hence, leading to non-pairwise interactions between the inclusions. This is in line with the general arguments used in the equilibrium context of non-active colloids in a bath of small non-active depletants, where multiple overlaps between depletion layers (occurring when the depletion layers are thick enough relative to the inclusion size) are described as the mechanism underlying non-pairwise many-body interactions between the inclusions  \cite{Lekkerkerker:2011}. The lowest order of such multiple overlaps occurs in the case of three inclusions at equal surface-to-surface distances in an equilateral-triangle formation  (Fig. 3.8 in Ref. \cite{Lekkerkerker:2011}), which is also the case considered in our analysis of three-body forces in an active bath. Our results thus show that, while an active bath can mediate strong repulsive two-body forces between inclusions, it can also produce comparably strong three-body forces with a more complex profile, displaying distinct repulsive and attractive regimes. Our analyses also indicate that multiple overlaps may be used as a measure for the significance of non-pairwise many-body contributions to the lowest order in the present non-equilibrium context with active particles. This point, however, remains to be confirmed by more systematic arguments. 

In the case of {\em chiral} active bath particles, the spatial distribution of active particles and the effective two- and three-body interactions mediated by them between the inclusions are found to be strongly dependent on the particle chirality: The layered structure of active particles is largely suppressed at elevated chirality strengths, giving rise to strong depletion of active particles from the narrow intervening region between the inclusions, and also their partial depletion from relatively thick, circular, zones further away from the inclusions. Consequently, the repulsive interaction forces mediated between the inclusions are weakened and, eventually, and for both the two- and the three-body forces, turn to relatively long-ranged, active and chirality-induced, depletion-type attractions at large enough particle chirality strength.  

The results reported in our work signify the important role of non-equilibrium active-particle layering and the resulting many-body interactions in active mixtures and also provide insight into the role of chirality. While in the equilibrium (non-active) context, standard theories (such as the free-volume and/or scaled particle theories in the case of hard spheres \cite{Lekkerkerker:2011,Likos:2001}) can be used to determine the phase behavior of the system, systematic approaches to study many-body effects and the phase behavior in active systems have been developed only quite recently (see Ref. \cite{cates_review} for a recent review). The importance of contributions from multi-body interactions in active particle/colloid mixtures can, nevertheless, be determined by resorting to numerical simulations. An efficient strategy is to develop direct, quantitative comparisons between implicit-active-particle simulations (in which only colloidal inclusions are explicitly modeled and they are then assumed to interact through effective forces mediated by active particles, such as those reported here, up to a given order in multi-body interactions) and explicit-active-particle simulations (in which both active particles and colloidal inclusions are modeled explicitly) \cite{Zaeifi:2017}. 

On the other hand, with recent progress in design and manipulation of synthetic Janus particles \cite{Walther:2013,Walther:softmatt2008,Jiang:2010,Perro:2005,sanojanus,valadares2010catalytic,ramin07,ramin_molmach,golestanian_njp,Douglass:NatPhot2012,Volpe:SciRep2014,Bianchi:SciRep2016,Buttinoni:PRL2013,Mano:JACS2005}, it is possible to develop active particles with different specifics (such as size, shape, and chirality); hence, exploring the non-equilibrium phase behavior of active particle/colloid mixtures by including these and other factors (such as inter-particle hydrodynamic coupling; see, e.g., Refs. \cite{BaskaranPNAS2009,EvansPhysF2011,MolinaSM2013,StarkPRL2014,wallattraction,wallattraction2,wallattraction3,ardekani,catesupstream,stark-wall,elgeti2009,gompper-wall,Hernandez-Ortiz1,elgeti2016}) appear as interesting potential directions for future research, where the question of effective interactions mediated by active constituents between inclusions is expected to play a central role. 

Other interesting problems that can be studied within the present context include the role of active noise (represented by fluctuating self-propulsion and angular velocities) in the dynamics of active particles \cite{Romanczuk:EPJ2012,Romanczuk:PRL2011,Grosmann:NJPhys2013,Grosmann:NJPhys2012,Romanczuk:EPJ2015}; these effects can result in features distinct from those obtained with passive, thermal, noise considered here.

%SECTION%%%%%%%%%%%%%%%%%%%%%%%%%%%%%%%%%%%%%%%%%%%%%%%%%%%%%%
%\section{Conflicts of interest}

%There are no conflicts of interest to declare. 

%SECTION%%%%%%%%%%%%%%%%%%%%%%%%%%%%%%%%%%%%%%%%%%%%%%%%%%%%%%
\section{Acknowledgements}
 
We thank the High Performance Computing Center (School of Computer Science, IPM) for computation time and Iran Science Elites Federation (ISEF) for partial support. A.N. acknowledges partial (Regular Associate) support from The Abdus Salam International Centre for Theoretical Physics (Trieste, Italy). We acknowledge useful discussions with M. Sebtosheikh and T. Jamali.

\bibliography{ref}

\end{document}